\documentclass{aa}
\usepackage{graphics}

\begin{document}

   \thesaurus{06     
              (13.07.1)}  
   \title{Are bright gamma-ray bursts a fair sample ?}

   \author{J-L. Atteia \inst{1}}

   \offprints{J-L. Atteia}

   \institute{ Centre d'Etude Spatiale des Rayonnements, CNRS/UPS,
                BP 4346, 31028 Toulouse Cedex 4, FRANCE\\
              email: atteia@cesr.fr
             }

   \date{Received 15 October 1999; accepted }

\titlerunning{}

   \maketitle

   \begin{abstract}

We conjecture that bright gamma-ray bursts (GRBs) are 
bright because they come from sources which are intrinsically 
over luminous and not because they come from nearby sources. 
We show that this hypothesis is supported by theoretical and observational 
arguments and that it explains some well-known properties of GRBs 
such as their Hardness-Intensity Correlation or the No-Host problem.
We discuss the consequences of this hypothesis on our understanding 
of the properties of the GRB population.

      \keywords{Gamma rays: bursts}
   \end{abstract}

%

\section{Introduction}

During the 90's the observations of the Burst and Transient Source Experiment 
(BATSE) on board the Compton Gamma-Ray Observatory provided a wealth of data 
on the properties of gamma-ray bursts at soft gamma-ray energies. 
The interpretation of these data was however complicated by our lack of knowledge 
of GRB distances.
This situation changed dramatically in 1997 with the discovery of afterglows 
at X-ray wavelength by BeppoSAX, which led to the discovery of visible afterglows 
and to the first distance determinations.
In this paper we show that the availability of burster distances sheds a new light 
on the interpretation of GRB properties measured at $\gamma$-ray energies.

The redshifts measured since 1997 (Table 1) have exposed the very broad 
dispersion of GRBs in luminosity.
With these new observations in mind, we discuss here the possibility 
that it is the burster intrinsic luminosity, and not the distance 
to the source, which determines the burst brightness measured at the earth. 
In Section 2, we show that this hypothesis is supported by the distribution
of GRB luminosities presently available. 
In Section 3, we explain that it also naturally explains
some well known (statistical) properties of the gamma-ray bursts.  
The consequences of this hypothesis on our understanding of
the GRB population are discussed in Section 4.

We now define our use of the words brightness and luminosity. 
We call the burst intensity measured at the earth brightness. 
The most common measures of brightness are the peak flux (in units of 
ph cm$^{-2}$ s$^{-1}$) and the fluence (in units of erg cm$^{-2}$).
We call the burst energy emitted at the source luminosity. 
The most common measures of luminosity are the peak luminosity 
(in units of ph s$^{-1}$) and the total luminosity (in units of erg).
In the absence of information on the beaming factor of the gamma-ray
emission, the peak and total luminosities are computed under the assumption
that the source is radiating isotropically; if the $\gamma$ emission 
is beamed toward us, the total energy radiated 
by the source could be much smaller.
In order to keep this paper simple we deal with a single measure of the
burst brightness (the fluence) and the corresponding measure of 
luminosity (the total luminosity).
We have checked that the use of the peak flux does not change our conclusions.

   \begin{table*}
      \caption[]{Luminosity of GRBs with known distances. 
      The luminosities have been computed for a standard universe with 
      Ho=70 km s$^{-1}$ Mpc$^{-1}$ and GRBs with a spectral index of -2.
      GB980425 has been excluded from this table since its association 
      with the supernova 1998bw at a redshift of 0.0085 remains controversial.}
         \label{lumi}
      \[
         \begin{array}{llll}
            \hline
            \noalign{\smallskip}
\mathrm{Name}  &  \mathrm{~~Fluence}  &  \mathrm{~~Redshift}  &  \mathrm{~~Total~Energy} \\
\noalign{\smallskip}
\hline
\noalign{\smallskip}
\mathrm{ GB970228} & ~~1.1~10^{-5} \mathrm{~(Hurley~et~al.~1997)} & ~~0.695 \mathrm{~(Djorgovski~et~al.~1999)} & ~~8.8~10^{51} \\
\mathrm{ GB970508} & ~~4.0~10^{-6} \mathrm{~(BATSE~Current~GRB~Catalog)} & ~~0.835 \mathrm{~(Metzger~et~al.~1997a,b)} & ~~4.4~10^{51} \\
\mathrm{ GB971214} & ~~1.3~10^{-5} \mathrm{~(BATSE~Current~GRB~Catalog)} & ~~3.42~ \mathrm{~(Kulkarni~et~al.~1998)} & ~~1.3~10^{53} \\
\mathrm{ GB980613} & ~~1.7~10^{-6} \mathrm{~(Woods~et~al.~1998)} & ~~1.096 \mathrm{~(Djorgovski~et~al.~1998b)} & ~~3.0~10^{51} \\
\mathrm{ GB980703} & ~~6.2~10^{-5} \mathrm{~(BATSE~Current~GRB~Catalog)} & ~~0.966 \mathrm{~(Djorgovski~et~al.~1998a)} & ~~8.8~10^{52} \\
\mathrm{ GB990123} & ~~5.1~10^{-4} \mathrm{~(BATSE~Current~GRB~Catalog)} & ~~1.60~  \mathrm{~(Kelson~et~al.~1999)} & ~~1.7~10^{54} \\
\mathrm{ GB990510} & ~~2.6~10^{-5} \mathrm{~(Kippen~et~al.~1999)} & ~~1.619 \mathrm{~(Vreesvijk~et~al.1999)} & ~~8.6~10^{52} \\
            \noalign{\smallskip}
            \hline
         \end{array}
      \]
   \end{table*}

\section{The Brightness Luminosity Correlation of GRBs}

The first measures of GRB redshifts have exposed the broad range
of intrinsic luminosities of these sources and their comparatively small 
range of distances. 
In order to provide a more quantitative view of this statement, 
we show in Table 2 various estimates of the dispersion of GRBs 
in distance and in luminosity.

The Table 2 strongly suggests that the parameter which primarily 
determines the burst brightness measured at the earth is 
not the distance of the source, but its intrinsic luminosity. 
This situation is the opposite of the standard candle hypothesis.
In the following we call it the Brightness Luminosity 
Correlation hypothesis (or BLUC).
Such a situation can only happen if the bursters have a particular
spatial distribution which is discussed in Section 4.3.

It is clear, however, that the number of redshifts 
which have been measured so far is too small to draw definite conclusions. 
A few tens of redshifts spanning the whole range of GRB brightnesses will probably 
be needed to transform what we still consider as a hypothesis into 
a firmly established GRB property.
Nevertheless we consider that, despite these uncertainties, the BLUC hypothesis 
has enough impact on our understanding of GRBs to deserve a discussion
of its consequences. 
This is done in the following sections.

\section{Brightness dependant properties of GRBs}

The brightness dependance of GRB properties has been extensively studied 
during the 90's as a way to unravel cosmological effects 
(e.g. spectral redshift or time dilation). 
The rationale behind this work was the concept that faint GRBs 
were more distant on average, and that they should consequently 
be more affected by the expansion of the universe.
These studies have disclosed two important properties of GRBs, 
the so-called Hardness-Intensity Correlation (or HIC) and the Time 
Dilation (TD). 

The BLUC conjecture, on the contrary, states that faint GRBs are 
not due to more distant sources but to sources
which are intrinsically less luminous.
This leads to a different interpretation of the Hardness-Intensity Correlation 
and of the Time Dilation which we discuss now.

\subsection{The Hardness-Intensity Correlation}
The Hardness-Intensity correlation is the observation that bright GRBs 
have on average harder energy spectra than faint GRBs. 
This property has been discussed by several authors in various contexts 
(e.g. Mallozzi et al. 1995; Dezalay et al. 1997 and ref. therein).
Within the context of BLUC, the Hardness-Intensity Correlation
simply reflects an underlying correlation between the luminosity 
of the source and its spectral hardness.
This effect is indeed expected within the framework of cosmological models
which invoke a plasma expanding at ultra-relativistic velocities, 
with a Lorentz factor ($\Gamma$) of several hundred.
The relativistic expansion of the emitting plasma multiplies the energy 
of the photons by a factor $\Gamma$ while it increases
the source luminosity by a huge factor (of the order of $\Gamma ^3$).
The combination of these two effects naturally produces a correlation between
the average photon energy and the luminosity of the source;
if the burst brightness reflects the radiated luminosity
(as postulated by the BLUC conjecture) this correlation 
is observed as HIC.

\subsection{The Time Dilation}
Time Dilation is the observation that
the timescales in the time histories of faint bursts are
typically longer than those measured in bright GRBs.
The reality of this effect and its interpretation have been subject
to ample discussions 
(e.g. Lestrade et al. 1993, Norris et al. 1994, Band 1994, Mitrofanov et al. 1996, 
Lee and Petrossian 1997, Stern et al. 1997 and ref. therein).
In the context of the BLUC hypothesis, TD means that the
timescales are longer in the light curves of intrinsically
subluminous bursts.

In the absence of a detailed model of the GRB prompt emission
there is no straightforward interpretation of this feature 
(unlike for the Hardness-Intensity Correlation). 
We note, however, that Ramirez-Ruiz \& Fenimore (1999) have recently found
that the faint peaks within a gamma-ray burst last longer 
than the more intense ones. 
This feature seems to supports the fact that low luminosity emission 
has longer characteristic timescales.

\subsection{The No-Host problem}
Another property of GRBs which has been discussed over the last years 
is the so-called No-Host problem, which is based on deep observations 
of the error boxes of several bright historical GRBs.
A detailed analysis of these error boxes 
(obtained by triangulation over the last 30 years)
shows that they do not contain bright galaxies.
If we assume that GRBs are hosted by normal galaxies, 
the apparent magnitude of 
the brightest galaxy in each error box can be used to derive 
a lower limit on the typical distance scale of those bright GRBs.
The no-host problem arises when one tries to extrapolate the distance 
derived for the brightest events to the population of faint GRBs.  
Schaefer (1999) shows that if faint GRBs are a distant version of 
bright bursts and if they are hosted by normal galaxies, 
they must be placed at very large distances (z $\approx$ 6).
An alternative explanation proposed by Schaefer, is that
GRBs do not reside in normal host galaxies.

The BLUC conjecture offers a third way to solve this problem.
If the brightness of GRBs is dominated by their intrinsic luminosity, 
faint GRBs are not distant versions of the brightest events.
They are instead bursts which are intrinsically less luminous 
but which have essentially the same distance scale (see below).
The BLUC hypothesis thus allows all GRBs to reside in normal galaxies.

   \begin{table}
      \caption[]{Contribution to the brightness dispersion of GRBs from 
      their intrinsic luminosity function and from their spread in distance.
      These numbers are computed for a standard universe with 
      Ho=70 km s$^{-1}$ Mpc$^{-1}$ and GRBs with a spectral index of -2.
      The standard deviation is given for the logarithm of the quantity.}
         \label{spread}
      \[
         \begin{array}{lll}
            \hline
            \noalign{\smallskip}
\mathrm{~~Parameter}  &  \mathrm{~~Luminosity}  &  \mathrm{~~Distance} \\
\noalign{\smallskip}
\hline
\noalign{\smallskip}
\mathrm{ ~~Dynamic~Range} & ~~560. &  ~~13.3 \\
\mathrm{ ~~Standard~Deviation} & ~~0.98 &  ~~0.38 \\
\mathrm{ ~~Correlation~with~brightness} & ~~0.93 &  ~~0.24  \\
            \noalign{\smallskip}
            \hline
         \end{array}
      \]
   \end{table}

\section{Discussion}

This section is devoted to a brief analysis of
the consequences that BLUC would have
on our understanding of the GRB population if
future redshift measurements confirm it.

\subsection{What is an average GRB ?}
As emphasized in the title, the BLUC hypothesis implies that 
bright GRBs are not representative of the bulk of the population.
They are intrinsically more luminous, with harder spectra and
cannot be used to infer the properties of average GRBs.
It seems thus better to use faint or intermediate GRBs
to derive the typical characteristics of the population 
(duration, energy of the peak of the SED...).

\subsection{The interpretation of the curve log(N)-log(S)}
Within the framework of BLUC the power law distribution of bright GRBs 
is not the consequence of the spatial distribution of nearby sources
but a direct measure of the luminosity distribution of gamma-ray bursts.
In the internal shocks paradigm, this distribution is closely related
to the distribution of the Lorentz factors of the emitting plasma. 
In this context it looks like an interesting coincidence that this slope
equals $-3/2$ which is precisely the value expected for sources homogeneously
distributed in a Euclidean space.

The break in the intensity distribution occurs when the luminosity function
is fully sampled for nearby bursters.
The interpretation of the curve Log(N)-Log(S) in the context of BLUC 
presents many other interesting properties which we plan 
to discuss in a future paper (Atteia et al., in preparation).
In a more general way, BLUC provides a natural explanation
of the fact that burst subclasses appear to have different intensity
distribution (e.g. Belli 1997, Pendleton et al. 1998, Tavani 1998).
Since the brightness distribution reflects the luminosity
distribution, it is not surprising that GRB subclasses
selected according to their temporal or spectral properties display
different luminosity (hence brightness) distributions.

\subsection{The GRB distribution in distance}
If the BLUC conjecture is correct, the distance of a GRB has little impact
on its observed brightness.
The only way to achieve such a situation is to consider bursters
which are restricted to a {\it limited range of distances}. 
This means that the bulk of the burster population occupies 
a shell-like volume around us with the more distant GRBs being 
only a few times farther than the nearby ones (while sources 
which are simply bounded
in space which can have a very broad {\it range} of distances). 
This seems to indicate that most GRBs occured at a particular epoch
of the life of the universe. In the context of the current ideas 
on the origin of GRBs, which relate them to violent stellar explosions, 
the BLUC conjecture thus appears compatible with the existence of a relatively 
well defined period of enhanced stellar formation. 

Another way to express this situation is to say that GRBs 
belonging to different classes of brightness have essentially 
the same distribution in distance.
An amusing consequence is that modest
GRB detectors (like PVO or ULYSSES) do sample the 
whole volume containing the GRBs, but for the brighest ones only. 
More importantly, this formulation provides an effective way 
to check the BLUC hypothesis via its prediction
that faint and bright bursts must have the same range of redshifts.
The availability of a few tens of redshifts in the next few years
with BeppoSAX and HETE-2 should confirm or discard this conjecture.
Should BLUC be confirmed, the redshifts already measured provide a good
idea of the extent of the GRB distribution in distance.

\begin{acknowledgements}
The author thanks J-P. Lestrade and R. Mochkovitch for valuable comments.
The author is also grateful to the BATSE team for making 
the Current BATSE GRB Catalog available at 
http://www.batse.msfc.nasa.gov/batse/grb/catalog/current/. 
\end{acknowledgements}

\end{document}